\documentclass[12pt]{article}
\usepackage{amsmath,amssymb,amsfonts,epsfig,graphicx,euscript}
\newcommand{\be}{\begin{equation}}
\newcommand{\ee}{\end{equation}}

\newcommand{\bse}{\begin{subequations}}
\newcommand{\ese}{\end{subequations}}
\newcommand{\bea}{\begin{eqnarray}}
\newcommand{\eea}{\end{eqnarray}}
\newcommand{\ba}{\begin{array}}
\newcommand{\ea}{\end{array}}

\makeatletter \@addtoreset{equation}{section}

\makeatother

\usepackage{cite}
\begin{document}
\baselineskip 18pt%

\begin{titlepage}
\begin{center}
\centerline{\Large{\bf{Drag force in asymptotically Lifshitz spacetimes  }}}%
\end{center}
\vspace*{5mm}
\begin{center}
{\bf K. Bitaghsir Fadafan}%
\vspace*{0.4cm}

{\it {Physics Department, Shahrood University of Technology,\\
P.O.Box 3619995161, Shahrood, Iran }}\\
{E-mails: {\tt bitaghsir@shahroodut.ac.ir}}%
\vspace*{1.5cm}
\end{center}

\begin{abstract}
We calculated drag force for asymptotically Lifshitz space times in
(d + 2)-dimensions with arbitrary dynamical exponent $z$. We find
that at zero and finite temperature the drag force has a non-zero
value. Using the drag force calculations, we investigate the DC
conductivity of strange metals.
\end{abstract}
\end{titlepage}

\section{Introduction}
A new method for studying different aspects of strongly coupled
quantum field theories is the AdS/CFT correspondence
\cite{Maldacena:1997re,Gubser:1998bc,Witten:1998qj, Witten:1998zw}
which has yielded many important insights into the dynamics of
strongly coupled field theories. Recently the application of this
duality in condensed matter physics called $AdS/CMT$ has been
studied \cite{Hartnoll:2009sz}. This duality is very useful to study
certain strongly coupled systems in $CMT$ by holography techniques
and to understand better their properties.

Methods based on $AdS/CFT$ relate gravity in $AdS_{d+2}$ space to
the conformal field theory on the $(d+1)$-dimensional boundary.
These conformal field
theories are invariant under the following scaling transformation%
 \be  (t,\vec{x})\rightarrow (\lambda \,t,\lambda\,\vec{x} ), \label{utranform}\ee%
However, in many condensed matter systems there are field theories
with anisotropic scaling symmetry. This unconventional scaling can
be illustrated as%
\be (t,\vec{x})\rightarrow (\lambda^z \,t,\lambda\,\vec{x} ),
\label{scaling} \ee%
where $z$ is the dynamical exponent. These field theories exist near
a critical phenomena and describe multicritical points in certain
magnetic materials and liquid crystals. In the case of $z=1$, theory
benefits from relativistic scale invariance. For $z=2$, there is a
2+1 dimensional field theory that so-called Lifshitz field theory.
This theory has a line of fixed points parameterized by $\kappa$ and
the lagrangian density is given by%
\be \mathcal{L}=\int\,dx^2\,dt\left( \left(\partial_t\phi\right)^2-
\kappa\,\left(\nabla^2\phi\right)^2 \right). \label{actionLif}\ee %
These fixed points are strongly coupled and they appear in strongly
correlated electrons in zero and finite temperature lattice models
\cite{Sachdev}. Another important theory with unconventional scaling
in (\ref{scaling}) is the theory whose symmetry group is
$Schr\ddot{o}dinger$ group, $Sch(d-1)$ and the geometry is given by
a deformation of AdS geometry \cite{shrodinger}. Many of the most
interesting examples in the condensed matter theory arise in the
Lifshitz case.

In this paper we consider massive charge carriers which are
described by the flavor branes in the Lifshitz space times. As it
was discussed in \cite{Hartnoll:2009ns}, considering massive charge
carriers in this background is the case of interest in modeling of
strange metals.  Gravity dulas have been investigated in
\cite{Kachru:2008yh}.%

 To calculate the drag force, one should consider a probe brane in this background. Adding flavor branes and finite-density
holography have been studied in \cite{flavor}. These probe branes
are related to a quantum critical theory and we consider massive
charge carriers interacting with this theory. Notice that these
carriers do not backreact on the quantum critical system. In the
massive case, the flavor brane forms a cigar-like shape with its tip
at $r_0$ and charge carriers correspond to strings stretching from
the tip of the cigar down to the horizon. At finite temperature, one
should consider a black brane in the background at a finite radial
position $r_h$. As a result, charge carriers on the flavor brane
correspond to the stretching strings from $r_0$ to the horizon. To
calculate the drag force, one should consider a stretching string
from the flavor brane and using prescription in
\cite{Herzog:2006gh,Gubser:2006bz,Herzog:2006se} find energy loss at
zero and finite temperature cases.

The most useful application of drag calculations in Lifshitz
background has been done in \cite{Hartnoll:2009ns}. They study
phenomenology of `strange metalls' and compute the electrical
conductivity. We discuss this application of drag force in the last
section. Using the drag force calculations, we investigate the DC
conductivity of strange metals and derive some results of
\cite{Hartnoll:2009ns}.\\

This paper is organized as follows. In the next section, we use the
proposed solutions in \cite{newmodel,R2} and discuss the energy loss
of massive charge carriers. We find that they lose energy even at
zero temperature. We compare this result with energy loss of
particle computed in the case of non-relativistic gravity dual to
field theory with Shrodinger CFT symmetry \cite{Akhavan:2008ep}.
Also in this case moving particle loses energy at zero temperature.
In section three, we consider the Lifshitz background embedded into
string theory \cite{stringdual}. We also find here a non-zero drag
force at zero temperature. In the last section we discuss computing DC conductivity from drag calculations.\\%

While this paper was in the final stages of preparation, it came to
our attention that the drag force in non-relativistic background
whose symmetry is $Schor\ddot{o}dinger$ group has been done in
\cite{Kluson:2009vy}. This calculation extends the results of
\cite{Akhavan:2008ep} and confirms a non-zero drag force at zero
temperature.

\section{Asymptotic Lifshitz space times}

In this section we provide backgrounds which are necessary for our
discussions. We consider non-relativistic holography in \cite{new
model}. In this study, Lifshitz geometry is a solution of gravity
coupled to a massive vector field. The (d+2)-dimensional spacetime action is%
\be S=\frac{1}{16\pi G_{d+2}}\int
d^{d+2}x\sqrt{-g}[R-2\Lambda-\frac{1}{2}\partial_{\mu}\phi\partial^{\mu}\phi
-\frac{1}{4}e^{\lambda\phi}F_{\mu\nu}F^{\mu\nu}]. \label{action}\ee %
where $\Lambda$ is the cosmological constant and massless scalar
field and abelian gauge field are matter fields of theory. The only
non-vanishing components of the field strength is $F_{rt}=q
e^{-\lambda \phi} r^{z-d-1}$ and $q$ is related to the charge of the
black hole. Based on this action, one finds the asymptotically
Lifshitz solution at zero temperature%

\bea
&&ds^{2}=L^{2}(-r^{2z}dt^{2}+\frac{dr^{2}}{r^{2}}+r^{2}\sum\limits^{d}_{i=1}dx^{2}_{i}),\nonumber\\
&&F_{rt}=qe^{-\lambda\phi}r^{z-d-1},~~~e^{\lambda\phi}=r^{\lambda\sqrt{2(z-1)d}},\nonumber\\
&&\lambda^{2}=\frac{2d}{z-1},~~~q^{2}=2L^{2}(z-1)(z+d),\nonumber\\
&&\Lambda=-\frac{(z+d-1)(z+d)}{2L^{2}}. \label{zerob}\eea%
In this solution, dilaton is not constant. However, exact solutions
can be found. To calculate the drag force, one should consider the
flavor probe branes in the background (\ref{zerob}) and study moving
massive charge carriers. From gauge-string duality, we consider a
trailing open string in the holographic direction. The action of
this open string is given by the Nambu-Goto action
\begin{eqnarray}
S=-T_0\int d\tau d\sigma\sqrt{-g }.\label{Nambo}
\end{eqnarray}
where $T_0$ is the tension of the string. The coordinates $(\sigma,
\tau)$ parameterize the induced metric $g_{ab}$ on the string
world-sheet and $g$ is the determinant of the world-sheet metric
$g_{ab}$
\begin{eqnarray}
-g=-detg_{ab }=(\dot X \cdot X')^2 - (X')^2(\dot X)^2,
\end{eqnarray}
where $X^\mu(\sigma, \tau)$ is a map from the string world-sheet
into space-time, and we define $\dot X = \partial_\tau X$, $X' =
\partial_\sigma X$, and $V \cdot W = V^\mu W^\nu G_{\mu\nu}$ where
$G_{\mu\nu}$ is the metric. The lagrangian density is given by
$\mathcal{L}=-T_{0}\sqrt{-g}$. The string equation of motion is
obtained as
\begin{equation}
\partial_{\rho}(\frac{\partial \mathcal{L}}{\partial x'})+\partial_t(\frac{\partial
\mathcal{L}}{\partial \dot{x}})=0.\label{eom}
\end{equation}
One has to calculate the canonical momentum densities $\pi^0_x,
\pi^0_t$ to find the total energy and momentum of the moving
particle in non-relativistic field theory
\begin{equation}
E=-\int^{\rho_0}_{\rho_h}d\rho\,\pi^0_t,
\,\,\,\,\,\,P=-\int^{\rho_0}_{\rho_h}d\rho\,\pi^0_x.
\end{equation}

\subsection{Drag force at zero temperature}

Now, we calculate drag force at zero temperature. We consider a
moving heavy point particle on the probe flavor brane in
d-dimensional space with the following ansatz %
\be
t=\tau,~~~~~~r=\sigma,~~~~~~x_1=x=v\,t+\xi(r),~~~~x_i=0(i\neq1),\label{ansatz}
\ee%
One finds from the equations of motion (\ref{eom}) that%
\be  \xi'^2=\frac{C^2\left( r^{2z-2}-v^2 \right)}{r^{2z+2}\left(
r^{2z+2}-C^2 \right)},\label{xit0} \ee
where $C$ is the constant of motion. The drag force that experiences by moving particle is%
\be F_{drag}=-T_o v^{2\left(\frac{z+1}{z-1}\right)}. \ee%
The drag force is independent of dimension of space. In (\ref{xit0}
), we considered string from boundary to infinity, then one finds
from numerator that there is no bound on the velocity and it can
change from zero to infinity. This is because of the fact that dual
theory is non-relativistic. This is an interesting result because
even though the system is at zero temperature the moving particle
losses its energy. We consider mass and momentum of particle as $M$
and $P$, respectively. Then $P=Mv$ and in the case of constant
momentum, the drag force will be found as $F_{drag}=\mu M v$. In
this way, one finds friction term as
$\mu=\frac{T_0}{M}v^{\left(\frac{z+3}{z-1}\right)}$.\\%

The Author in \cite{Akhavan:2008ep} found a non-zero drag force at
zero temperature. They studied non-relativistic three dimensional
CFT at zero and finite temperature. They found that unlike the AdS
case where one only gets a casual speed limit, in the
non-relativistic case one arrives no speed limit and non-zero drag
force. We also find same results in asymptotically Lifshitz
spacetimes with arbitrary critical exponent. As a result the
non-zero drag force would be considered as a common properties of
non-relativistic spacetimes.

\subsection{Drag force at finite temperature}

Unlike the case of $Schr\ddot{o}dinger$ conformal group, it is
difficult to obtain analytic black hole solutions in Lifshitz
spacetimes. Actually, the problem of finding analytic exact black
hole solutions with asymptotically Lifshitz geometry turned out to
be a highly non-trivial problem. However, there are known solutions.
For example, black hole solutions with z = 2 in four dimensions were
studied in \cite{Numeric} and black holes in asymptotically Lifshitz
spacetimes with arbitrary critical exponent were investigated in
\cite{z}. Topological black holes and other solutions were proposed
in \cite{topological,new model}. For other recent solutions on
Lifshitz black holes see \cite{others}.

We consider the black brane solution of (\ref{action}) in asymptotic
Lifshitz (2+1)-dimensional spacetime proposed in \cite{new model}
\be ds^2=L^2\left( -r^{2z}\,f(r) dt^2+\frac{dr^2}{r^2\,f(r)}+r^2
d\vec{x}_i^{\,2} \right),\,\,\,\,\,\,
f(r)=1-\left(\frac{r_+}{r}\right)^{z+d}. \ee%
The matter fields are a massless scalar and an abelian gauge field
and the other fields remain the same as those in the zero
temperature case.
The temperature is %
\be T_H=\frac{(z+d) r_+^z}{4 \pi}, \ee%
and the black hole entropy is given by
\be S_{B}=\frac{V_d L^d r_+^d}{4 G_{d+2}}, \ee%
which $V_d$ is the volume of the $d$ dimensional spatial coordinates
$\vec{x}_i$.\\%

We consider the moving heavy point particle in $x$ direction and
following the ansatz in (\ref{ansatz}), one finds that%
\be \xi'^2=\frac{C^2 \left( r^{2z-2}-\frac{v^2}{f(r)} \right)}{f(r)
r^{2z+2}\left(L^4 f(r) r^{2z+2}-C^2 \right)}, \label{xesi1}\ee%
where $C$ is constant of motion. From reality condition of above
equation, one finds that $C=v L^2 r_c^2$ and $r_c$ is the root of
this equation%
\be \left( 1-\left(\frac{r_+}{r_c}\right)^{z+d} \right)
r_c^{2z-2}-v^2=0, \label{root1}\ee %
The drag force is given by %
\be F_{drag}=-T_0\, v L^2 r_c^2.\ee%
We consider some special cases and derive the drag force as the
following
\begin{itemize}
\item{$d=2,z=1$\\
In this case, there are two dimensional spatial coordinates and
$z=1$ means the isotropic theory. The drag force in terms of the
temperature of the field theory, $T_H=\frac{3 r_+}{4
\pi}$ }, is given by %
\be F_{drag}=-T_0 L^2 \left(\frac{4 \pi}{3}\right)^{\frac{2}{3}}
T_H^{\frac{2}{3}} \frac{v}{\left(1-v^2\right)^{\frac{2}{3}}}. \ee

\item{$d=2,z=2$\\
This specific example is known as the Lifshitz model. This case
appears in systems of strongly correlated electrons in condensed
matter physics. Having a holographic description for these phenomena
would be of great importance to investigate some properties of
strongly coupled systems in condensed matter. One finds drag force
as \be F_{drag}=-T_0 L^2 v^2 \left(
\frac{v^2}{2}-\sqrt{\frac{v^2}{2}+\pi^2\,T_H^2} \right). \ee
  }
\end{itemize}%
It is clear that in each case the temperature dependency of drag
force is different. It is straightforward to discuss drag force in
other cases with spatial dimensions more than $d=2$ and different
values for $z$.

\subsection{$R^2$ corrections to the drag force}

Now we study $R^2$ corrections. These corrections to
five-dimensional asymptotically Lifshitz spacetimes have been
studied in \cite{R2}. The specific example is Gauss- Bonnet model.
Black brane solution has been found perturbatively as

\begin{equation}
ds^{2}=L^{2}[-g(u)(1-u)dt^{2}+\frac{1}{h(u)(1-u)}du^{2}+\frac{r_{+}^{2}}{u^{A}}
(dx^{2}_{1}+dx^{2}_{2}+dx^{2}_{3})],
\end{equation}
where
\begin{eqnarray}
& &g(u)=r^{2z}_{+}u^{-\frac{4z}{z_{0}+3}}(1+u)\left(1+\lambda_{\rm
GB}(1-u^{2})\right)\exp[4\lambda_{\rm
GB}\frac{z_{0}-1}{z_{0}+3}u^{2}],\nonumber\\
& &h(u)=\frac{1}{4}(z_{0}+3)^{2}u^{2}(1+u)\left(1+\lambda_{\rm
GB}(1-u^{2})\right). \label{R2metric}\end{eqnarray}%
where $\lambda_{GB}$ is the Gauss-Bonnet coupling constant.
and%
\be u^2=(\frac{r_{+}}{r})^{z_{0}+3}~~~A=\frac{4}{z_{0}+3},~~~z=z_0+2 \lambda_{GB}(z_0-1).\ee %
The horizon of the black brane locates at $u=1$ and the boundary
locates at $u=0$. the Hawking temperature is%
\be T_H=\frac{(z_0+3)r_+^z}{4 \pi}\left( 1+2 \lambda_{GB}\left(
\frac{z_0-1}{z_0+3}\right) \right).\label{TR2} \ee%
The drag force in the case of $z_0=1$ has been calculated in
\cite{Fadafan:2008gb}.\\

We consider the moving heavy point particle in $x$ direction and
following the ansatz in (\ref{ansatz}), one finds that%
\be \xi'^2=\frac{C^2\left( \frac{g(u)}{h(u)}-\frac{r_+^2\,v^2}{u^A
h(u)
(1-u)}\right)}{\frac{r_+^2(1-u)g(u)}{u^A}\left(\frac{L^4\,r_+^2(1-u)g(u)}{u^A}-C^2\right)
}, \label{xesiR2}\ee%
From reality condition of above
equation, We should find roots of this equation%
\be g(u_c)-\frac{r_+^2\,v^2}{u_c^A (1-u_c)}=0, \ee%
Regarding this equation, drag force can be found as%
\be F_{drag}= -T_0 r_+^2 v\left(\frac{1+u_c}{u_c^A}\right). \ee%
Notice that dependency of drag force to the temperature of the field
theory is complicated.\\%

One can study charge effects on the drag force, too. The charged
Lifshitz black hole solutions in general (d + 2)- dimensions have
been investigated in \cite{Charge}. Also the Gauss-Bonnet
corrections to such black holes in five dimensions have been
calculated perturbatively. Using these solutions, one can study drag
force in these backgrounds.

\section{Drag force in String Duals of Non-relativistic Lifshitz-like Theories}

The aim of this section is to study some features of
(1+2)-dimensional non-relativistic field theory using supergravity
solution in type IIB string theory. Since we are dealing with string
theory, it is natural to consider a semi-classical string in this
background. However, it is difficult to embed the Lifshitz
background into string theory. Some no-go theorems for string theory
duals of non-relativistic Lifshitz like theories have been proposed
in \cite{No-go}. They propose that classical solutions in type IIA
and eleven-dimensional supergravities are not possible. (These
solutions are expected to be dual to (2+1)- dimensional
Lifshitz-like theories.) Based on holographic constructions of
fractional quantum Hall effect (FQHE) via string theory, authors of
\cite{FGHE} proposed D3-D7 solutions. Using this construction, the
embedding of anisotropic background into type IIB string theory was
studied in \cite{string dual}. However, the scaling behavior in this
solution is different and the anisotropy of the scale transformation
is only through one of the three spatial directions. As a result, it
corresponds to a classical Lifshitz point. Also since it has a
non-constant dilaton, the anisotropic scale invariance only holds at
the leading order of interactions. In the context of AdS/CFT
correspondence an open string can be associated to a Wilson loop in
the dual field theory and one can consider particle at the end of
this semi-classical open string at the boundary. Regarding this
study we consider a moving point particle in a strongly correlated
system and calculate the drag force.

\subsection{Drag force at zero temperature}

We study drag force at zero temperature. Spacetime metric in
the Einstein frame is given by \cite{string dual}%
 \bea
ds^2_E=\tilde{R}^2\left[r^2(-dt^2+dx^2
+dy^2)+r^{\frac{4}{3}}dw^2+\frac{dr^2}{r^2}\right] +R^2ds_{X_5}^2,
\eea%
This metric is invariant under the scaling
\begin{equation}
(t,x,y,w,r)\to \left(\lambda t,\lambda x, \lambda y,
\lambda^{\frac{2}{3}} w, \frac{r}{\lambda}\right),
\end{equation}
and therefore is expected to be holographically dual to
Lifshitz-like fixed points with space-like anisotropic scale
invariance. One can redefine the radius coordinate $\rho\equiv
r^{\frac{2}{3}}$ and rescale $(t,x,y,w)$ accordingly. Then metric in
string frame will be as the following
 \begin{equation}
  ds^2_E=\tilde{R}^2\left[\rho^3(-dt^2+dx^2
+dy^2)+\rho^{2}dw^2+\frac{d\rho^2}{\rho^2}\right] +R^2ds_{X_5}^2.
 \end{equation}
This can be regarded as gravity duals of Lifshitz-like fixed points
with $z=3/2$.\\%

Now we study a moving heavy particle in "x" and "w" directions. We
expect different behaviors in two directions. Because "w" direction
is an anisotropic direction but "x" is not.

\textbf{Moving in x direction, a time dependent solution: }\\%

In this case, we study moving particle in $x$ direction and consider
a time dependent solution as the following ansatz
\begin{eqnarray}
X^{\mu}=(t=\tau,\,\,\,x=vt+\xi(\rho),\,\,y=0,\,\,w=0,\,\,\rho).
\end{eqnarray}
one finds the Nambu-Goto action
\begin{eqnarray}
S=-\frac{1}{2 \pi \alpha'} \int dt d\rho \tilde{R}^2 \sqrt{\rho(1-
v^2)+\xi'^2 \rho^6}.
\end{eqnarray}
It would be straightforward to calculate $\xi'$ from the above
equation
\begin{eqnarray}
\xi'^2=\frac{A^2\rho (1-v^2)}{\rho^6 \left(\tilde{R}^4 \rho^6
-A^2\right)},
\end{eqnarray}
where $B$ is the constant of motion. By studying the reality
condition for lagrangian density $\mathcal{L}=-T_{0}\sqrt{-g}$ and
therefore for $\xi'^2$, one finds that the constant of motion can be
chosen arbitrary. In the special case of moving with speed of light,
one finds $F_{drag}=-T_0\,\tilde{R}^2\, \rho^3$. One observes that
there is a bound on the velocity which it was expected because $x$
direction is not anisotropic direction.\\

\textbf{Moving in $w$ direction, a time dependent solution }\\%

In this case we ask about the moving non-relativistic particle in
anisotropic direction. The ansatz is $w=vt+\xi(\rho)$ and from the
Nambu-Goto action (\ref{Nambo}), one finds that
\begin{eqnarray}
\xi'^2=\frac{B^2(\rho-v^2)}{\rho^5(\rho^5-B^2)},\label{Num}
\end{eqnarray}
$\xi'^2$ must be a real parameter and with this condition, $B$ can
be found. Also it is clear from the numerator of (\ref{Num}) that
there is no bound on the velocity of particle and it can be changed
from zero to infinity. This result is reasonable, because "w"
direction is anisotropic and the dual theory is non-relativistic. As
a result, one finds a non-zero drag force on a moving particle in
$w$ direction and at zero temperature as
\begin{equation}
F_{drag}=\frac{dp}{dt}=-T_0\,B=-T_0 v^5.\label{zero2}
\end{equation}
One can consider the momentum and mass of particle as $P$ and $M$,
respectively and use the non-relativistic relation $P=M v$. We
rewrite drag force in terms of $P$ and derive momentum of particle as%
\be P(t)=\left( \frac{4 T_0}{M^5} \right)^{\frac{1}{4}}
\,\frac{1}{t^{\frac{1}{4}}}.\ee%
It would be interesting to calculate friction coefficient of moving
particle. Using the relation $\dot{P}=-\mu M v$, one finds that
friction coefficient is velocity dependent, $\mu= \frac{T_0}{M}
v^4$. \\%

In this case although the system is at zero temperature, the moving
particle losses its energy. This phenomena is in common with
\cite{Akhavan:2008ep} which they consider
non-relativistic three dimensional CFT at zero temperature.\\%

\subsection{Drag force at finite temperature}

It was shown that an $AdS$ space with a black brane is dual to
conformal field theory at finite temperature \cite{Witten:1998zw}.
We use the extension of AdS/CFT correspondence in the case of an
anisotropic spacetime. This gravity dual is known as the string
theory duals Lifshitz-like fixed points \cite{string dual}. The
metric in the Einstein frame is
\begin{equation}
ds^2_E=\tilde{R}^2\left[r^2(-f(r)dt^2+dx^2
+dy^2)+r^{\frac{4}{3}}dw^2+\frac{dr^2}{r^2f(r)}\right]
+R^2ds_{X_5}^2, \label{bhsol}
\end{equation}
where
\begin{equation}
f(r)=1-\frac{\mu}{r^{\frac{11}{3}}}.\label{frnon}
\end{equation}
The constant $\mu$ represents the mass parameter of the black brane.
and the Hawking temperature is
\begin{equation}
T_H=\frac{11}{12\pi}\mu^{\frac{3}{11}}.
\end{equation}

We study a moving object in the hot gauge theory and in the bulk
space a moving open string should be considered, too.\\%
Based on the Nambu-Goto action of string in (\ref{Nambo}) and
equation of motion in (\ref{eom}) the simplest solution for the
equation of motion is $x=constant$. In this case the string is
stretched from the probe D-brane at $\rho=\rho_m$ to the horizon at
$\rho=\rho_h$, straightforwardly. In the other word we have a static
particle without any motion. The energy of particle in this case is
associated with the rest mass of  the particle which is obtained by  %
\be M_{rest}=\frac{T_0\,\tilde{R}^2}{2}\left( \rho_m^2 - \rho_h^2
\right). \label{massgap}\ee%
One can compare this result with the rest mass of a static quark in
$\mathcal{N}=4$ SYM theory \cite{Herzog:2006gh}. In our study, one
may use the relation $E_{gap}=MC^2$ and interpret (\ref{massgap}) as
the energy scale of bulk excitations at the position of flavor
brane.

\textbf{Moving in x direction, a time dependent solution}:\\%

We consider the moving particle in $x$ direction and consider a time
dependent ansatz $x=vt+\xi(r)$, with this choice, one finds that
\begin{eqnarray}
\xi'^2=\frac{1-\frac{v^2}{f(r)}}{r^4 f(r) \left(\tilde{R}^4 r^4
f(r)-C^2\right)},
\end{eqnarray}
from this equation, we can find the critical radius where numerator
and denominator change their sign
\begin{eqnarray}
r_c=\left(\frac{\mu}{1-v^2}\right)^\frac{3}{11},
\end{eqnarray}
and finally the drag force on the particle is given by
\begin{eqnarray}
\frac{dp}{dt}=-(\frac{12\,
\pi}{11})^2\,\tilde{R}^2\,T_H^2\,\frac{v}{(1-v^2)^\frac{6}{11}}.
\end{eqnarray}
And it is clear that the moving particle loses its energy at finite
temperature case, too.\\%

\textbf{Moving in w direction, a time dependent solution}:\\%

It would be interesting to study a moving object in "w" direction.
This is an anisotropic direction which spacetime violates lorentz
symmetry. One can consider the following ansatz
\begin{eqnarray}
X^{\mu}=(t=\tau,\,\,\,x=0,\,\,y=0,\,\,w=vt+\xi(r),\,\,r).
\end{eqnarray}

The constant of motion can be considered as $D$ and from equation of
motion, one finds that
\begin{eqnarray}
\xi'^2=\frac{D^2(1-\frac{v^2\,r^{\frac{-2}{3}}}{f(r)})}{r^{\frac{10}{3}}
f(r) \left(\tilde{R}^4 r^{\frac{10}{3}} f(r)-D^2\right)},
\end{eqnarray}
Both numerator and denominator must change sign at the same root,
and from numerator one should solve $r^{\frac{11}{3}}-v^2 r^3-\mu=0$
to find drag force. We name the root as $r_c$ and one finds drag force as%
\be F_{drag}=-T_0\, v\, r_c^{\frac{4}{3}}.\ee%
Based on this relation, one can discuss on the drag force, too.

\section{Discussion}

In this paper we have calculated drag force for asymptotically
Lifshitz space times in (d + 2)-dimensions with arbitrary dynamical
exponent $z$ from gauge-string duality. We have used the proposed
solutions in \cite{new model,R2}. The finite temperature behavior of
Wilson loops as an application to strongly coupled gauge theories in
3+1 dimensions has been studied in \cite{Numeric,Koroteev:2009qr}.
By analyzing action in (\ref{actionLif}), one concludes that the
boundary theory can be viewed as a gauge theory in 2+1 dimensions
with a dimensionless coupling constant and as a result the theory
perhaps has some features in common with conventional gauge theory
in 3 + 1 dimensions \cite{Numeric}. Having Wilson loops on the
gravity side, one can study drag force in the gauge theory side.
From gauge-string duality, one should consider a hanging string from
the boundary to the horizon. The end point of the string represents
the particle that is charged under the gauge field. We have
considered a moving
heavy point particle and calculated drag force at zero and finite temperature non-relativistic field theories.\\%

We have found the energy loss of moving heavy point particle. For
the zero temperature background, we found that particle loses energy
even though at zero temperature. Also we considered the string
theory dual to Lifshitz-like fixed points with anisotropic scale
invariance which proposed in \cite{string dual} and studied drag
force. We found a non-zero drag force in the case of zero
temperature, too. In this case, there are anisotropic and isotropic
directions. We have found a non-zero drag force when particle is
moving in these directions in (\ref{zero2}). Then, we compared our
results with drag force in the case of field theory whose symmetry
group is $Schr\ddot{o}dinger$ group in \cite{Akhavan:2008ep}. In
this reference, the energy loss of particle computed in the case of
non-relativistic gravity dual to field theory with
$Schr\ddot{o}dinger$ CFT symmetry. Also they found that a moving
particle loses energy at zero temperature. Based on these studies,
we conclude that this could be a common property of non-relativistic
field theories. Holographic description of strongly correlated
systems in condensed matter physics implies a non-zero drag force on
a moving heavy carrier at zero and finite temperature.

The most useful application of drag calculations in Lifshitz
background has been done in \cite{Hartnoll:2009ns}. They study
phenomenology of `strange metalls' and compute the electrical
conductivity. It would be interesting to relate our results to
calculation of DC conductivity.

In order to calculate DC conductivity, an electric field should be
turn on on the D-brane probe and the resultant current $ J^x $ can
be computed in the boundary \cite{Karch:2007pd}. The
conductivity $\sigma(E,T)$ is found from Ohm's low%
\be \sigma(E,T)=\sqrt{\sigma_o^2+\sigma^2}
 \ee%
where $\sigma_0$ is a constant term and arises from thermally
produced pairs of charge carriers. By increasing the mass of
carriers, $\sigma_0$ can be made arbitrary small and the leading
term in conductivity will be $\sigma$. Based on the results of
\cite{Karch:2007pd}, we discuss calculating of leading term in
conductivity, $\sigma$, by studying the properties of a moving
single string. The Authors in
\cite{Hartnoll:2009ns} found that%
\be \sigma^2=\left(\frac{2\pi
\alpha'}{L^2}\right)^2 r_*^{-4}\,(J^t)^2\label{sigma}\ee%
where $r_*$ is the root of the following equation%
\be r^{2z+2}\,f(r)-(2\pi \alpha')^2\,E^2=0.\label{Econdition}\ee%
An important result of \cite{Hartnoll:2009ns} is based on the
relation (\ref{sigma}). This equation exhibits the power-law for the
DC resistivity, $\rho\sim\frac{T^{\frac{2}{z}}}{J^t}$. As it was
discussed in \cite{Hartnoll:2009ns}, this behavior is generic in a
regime of dilute charge carriers which are coupled to a Lifshitz
matter. Now we derive this result from our drag force calculations.\\

We consider the quasi-particle description and write the equation of
motion for them at the equilibrium where the external force is
$f=E$. At large mass limit, only charge carriers contribute to
current and one may express $ J^x $ in terms of velocity of the
quasi-particles, $ J^x = J^t \,v $. Regarding Ohm's law(
$J^x=\sigma\,E$) one finds the leading term in conductivity as%
\be \sigma=\frac{v\, J^t}{E}. \ee%
Based on the drag force calculations at finite temperature and from
numerator of
(\ref{xesi1}), one finds velocity of the quasi-particle as%
\be v^2=r_*^{2z-2}\,f(r_*) \label{v1}\ee%
The drag force is related to the constant of motion $C$ which can be
found from denominator of (\ref{xesi1}). Also at equilibrium,
$C^2=(2\pi
\alpha')^2\,E^2$ then%
\be E^2=\frac{L^4\,r_*^{2z+2}\,f(r_*)}{(2\pi \alpha')^2}\label{E}\ee
From (\ref{v1}) and (\ref{E}), one derives conductivity as%
\be \sigma=\frac{2\pi
\alpha'}{L^2} r_*^{-2}  J^t  \label{cond1}\ee%
which is the same as (\ref{sigma}). When external field is very
weak, one concludes that $r_*\sim r_+$ and as a result $\sigma \sim
\,\frac{J^t }{T^{\frac{2}{z}}}$ which confirms the result of
\cite{Hartnoll:2009ns}.

It would be interesting to study $R^2$ corrections to the DC
conductivity for asymptotically Lifshitz backgrounds. One should
consider calculation of the drag force from (\ref{xesiR2}). It is
straightforward to find $\sigma_{R^2}=\frac{2\pi \alpha'}{L^2}
r_*^{-2}  J^t$ and for small electric field one concludes that
$r_*\sim r_+$ where $r_+$ is radial horizon and it is related to the
temperature of the matter (\ref{TR2}). As a result $R^2$ corrections
to conductivity is given by %
\be \sigma_{R^2}\sim\left(\frac{4\pi}{ z_0+3+2\lambda_{GB}(z_0-1)
}\right)^{-2/z}\,T_H^{-2/z}\,J^t. \ee%
From this equation, One can study the effect of Gauss-Bonet coupling
constant $\lambda_{GB}$ on the DC conductivity and resistivity. In
the case of $z_0=1$ one finds the result of \cite{Hartnoll:2009ns}.

Calculating DC conductivity at zero temperature is straightforward.
One should find the velocity of massive charge from numerator of
(\ref{xit0}) and based on the relation between velocity
and conductivity, one finds%
\be \sigma_{T=0}=\frac{2\pi \alpha'}{L^2} r_c^{-2}  J^t
.\ee%
where $r_c$ is not the same as $r_*$ in (\ref{cond1}).\\
It is interesting to investigate calculation of conductivity from
drag force in type IIB string theory which it was studied in
\cite{string dual}. We consider the finite temperature matter and
turn on an small electric field in "x" and "w" directions. In these
directions one finds that $\sigma_x\sim\frac{1}{T^2}$ and
$\sigma_w\sim\frac{1}{T^{4/3}}$. As we expected, conductivity is
different in these directions.

\section*{Acknowledgment}
I would like to thank M. Alishahhiha for very useful discussions and
specially thanks J. McGreevy for reading the manuscript and
comments.

\end{document}